\documentclass[apj]{emulateapj}
\usepackage{apjfonts}
\slugcomment{To appear in ApJ Letters}

\shorttitle{Radial Distribution of Star Formation at $z\sim 1$}
\shortauthors{Nelson et al.}

\begin{document}

\gdef\ha{H$\alpha$}
\gdef\ew{${\rm EW}({\rm H}\alpha)$}
\gdef\msun{M$_{\odot}$}

\title{The Radial Distribution of Star Formation in Galaxies at $z\sim 1$ From
the 3D-HST Survey}

\author{Erica June Nelson\altaffilmark{1}, 
Pieter G.\ van Dokkum\altaffilmark{1}, 
Ivelina Momcheva\altaffilmark{1},
Gabriel Brammer\altaffilmark{2},
Britt Lundgren\altaffilmark{3},
Rosalind E. Skelton\altaffilmark{1},
Katherine E. Whitaker\altaffilmark{4},
Elisabete Da Cunha\altaffilmark{5},
Natascha F\"orster Schreiber\altaffilmark{6},
Marijn Franx\altaffilmark{7},
Mattia Fumagalli\altaffilmark{7},
Mariska Kriek\altaffilmark{8},
Ivo Labbe\altaffilmark{7},
Joel Leja\altaffilmark{1},
Shannon Patel\altaffilmark{7},
Hans-Walter Rix\altaffilmark{5},
Kasper B.\ Schmidt\altaffilmark{9},
Arjen van der Wel\altaffilmark{5},
Stijn Wuyts\altaffilmark{6}}

\altaffiltext{1}{Astronomy Department, Yale University, 
New Haven, CT 06511, USA}
\altaffiltext{2}{European Southern Observatory, Alonson de C\'ordova
3107, Casilla 19001, Vitacura, Santiago, Chile}
\altaffiltext{3}{Department of Astronomy, University of Wisconsin-Madison, 
Madison, WI 53706, USA}
\altaffiltext{4}{Astrophysics Science Division, Goddard Space Flight Center,
Greenbelt, MD 20771, USA}
\altaffiltext{5}{Max Planck Institute for Astronomy (MPIA), K\"onigstuhl 17,
69117, Heidelberg, Germany}
\altaffiltext{6}{Max-Planck-Institut f\"ur extraterrestrische Physik,
Giessenbachstrasse, D-85748 Garching, Germany}
\altaffiltext{7}{Leiden Observatory, Leiden University, Leiden, The
Netherlands}
\altaffiltext{8}{Department of Astronomy, University of California,
Berkeley, CA 94720, USA}
\altaffiltext{9}{Department of Physics, University of California,
Santa Barbara, CA 93106, USA}


\begin{abstract}

The assembly of galaxies can be described by the distribution 
of their star formation as a function of cosmic time.  
Thanks to the WFC3 grism on HST it is now possible to 
measure this beyond the local Universe.
Here we present the spatial distribution of \ha\ emission for a 
sample of 54 strongly star-forming galaxies at 
$z\sim 1$ in the 3D-HST Treasury survey. 
By stacking the \ha\ emission we find that  star formation 
occurred in approximately exponential distributions at $z\sim 1$, 
with median S{\'e}rsic index of $n=1.0\pm0.2$. 
The stacks are elongated with median axis ratios of  $b/a=0.58\pm0.09$
 in \ha\, consistent with (possibly thick) disks at random orientation angles.
  Keck spectra obtained for a subset of eight of the galaxies show clear 
  evidence for rotation, with inclination corrected velocities of 90 to 330\,km/s. 
  The most straightforward interpretation of our results is that star formation in
   strongly star-forming galaxies at $z\sim 1$ generally occurred in disks. 
  The disks appear to be "scaled-up" versions of nearby spiral galaxies: 
  they have EW(\ha\,)$\sim$100\AA\, out to the solar orbit and they have star formation 
  surface densities above the threshold for driving galactic scale winds.
 
\end{abstract}

.

\section{Introduction}

Galaxy formation is a complex process, involving starbursts,
mergers, and strong gas flows 
(e.g., {Hopkins} {et~al.} 2006; Brooks {et~al.} 2009; {Dekel} {et~al.} 2009). 
Furthermore, stellar migration and secular processes can change the
structure of galaxies at late times 
(e.g., Ro{\v s}kar {et~al.} 2008; Grand, Kawata, \& Cropper 2012). 
Therefore, even if we could perfectly locate and age
 date every star in the Milky Way, 
  we still couldn't say where and with what 
  structural and kinematic properties those stars formed.
The only way to establish where a galaxy's stars formed,
  and hence how it was assembled, is 
  to map the  star formation \textit{while} those stars were forming.

Obtaining \ha\, and stellar continuum maps of $z\geq1$ galaxies,
 with the $\sim1$\,kpc resolution necessary to put constraints on the 
 spatial distribution of star formation, is challenging and has so far only been
 possible using a combination of adaptive optics and integral field units (IFUs)
 on 8--10\,m class telescopes.
These studies paint a complex picture: they find that star forming
galaxies at $z\sim 2$ are
a mix of  ``puffy'' and often clumpy rotating disks, 
 mergers, and more compact
dispersion-dominated objects
 (e.g., {Genzel} {et~al.} 2008; {Shapiro} {et~al.} 2008; Cresci {et~al.} 2009; {F{\"o}rster Schreiber} {et~al.} 2009; {Law} {et~al.} 2009; Jones {et~al.} 2010; Mancini {et~al.} 2011). 

Some studies claim that there are trends in the structural properties of \ha\
as a function of redshift implying that the way galaxies
assemble their stars varies fundamentally as a function of cosmic time.
In particular,
Epinat {et~al.} (2009) and Kassin {et~al.} (2012) suggest that galaxies become cooler and more
rotation-dominated with time, $z\sim1-0$ being the epoch of ``disk settling''.
This is interesting because most of the stars in 
the disks of galaxies like the Milky Way were
formed in this epoch.

It is now possible to obtain high spatial resolution ($\lesssim1$\,kpc)
information on \ha\, emission at $z\sim1$ with a high Strehl ratio, owing to 
the near-IR  slitless
spectroscopic capabilities provided by the WFC3 camera on
the Hubble Space Telescope (HST).
In {Nelson} {et~al.} (2012) we
used data taken as part of the 3D-HST survey to 
build on previous ground-based studies by mapping 
the  \ha\, and stellar continuum with high resolution 
for a sample of 57 galaxies at $z\sim1$ and 
 showed that star formation broadly follows the rest-frame optical light,
but is slightly more extended. 
Here we stack the \ha\ maps of these galaxies to construct high
signal-to-noise ratio (S/N) radial profiles and measure structural
parameters of stellar continuum emission and star formation.
These unique, high spatial resolution data from 3D-HST
are combined with kinematics from
 the Near Infrared Spectrometer (NIRSPEC) on the
W. M. Keck telescope ({McLean} {et~al.} 1998) 
We assume H$_0=70$ km s$^{-1}$Mpc$^{-1}$, $\Omega_M=0.3$, 
and $\Omega_{\Lambda}=0.7$.

\section{3D-HST And Sample Selection}

\begin{figure}
\centering 
\includegraphics[trim=10mm 40mm 0mm 40mm,clip,width=0.5\textwidth]{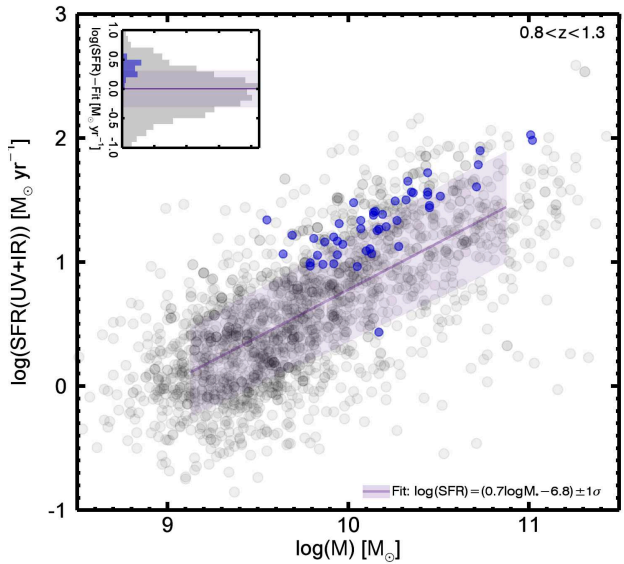}
\caption{
Location of selected galaxies in the SFR-Mass plane.
Six sources without IR photometry are not shown in the figure.
The histogram shows the position of the selected galaxies relative to 
the rest of the galaxies by dividing a power law fit out of the distribution.
The purple line shows the fit, the light purple region delineates $\pm1\sigma$.
The selected galaxies 
appear to lie largely on the upper half of the `main sequence' in SFR(IR+UV). 
\label{sel}}
\end{figure}

The 3D-HST survey, 
a 248-orbit Treasury program on the Hubble Space Telescope, 
supplies the 2D emission line maps needed to measure the 
spatial distribution of star formation.
The WFC3 G141 grism provides spatially resolved spectra of all
sources in the field.  The wavelength range of the G141 grism,
$1.15\mu{\rm m}<\lambda<1.65\mu$m, covers the
H$\alpha$ emission line for galaxies in the redshift range
$0.7<z<1.5$. The survey also provides
broad-band near-infrared imaging in the F140W filter, which samples
the rest-frame $R$ band for $z\sim1$.
Using the same camera under the same conditions
to map the both \ha\, and continuum emission allows 
us to compare their distributions directly, 
a crucial advantage of this strategy.
 The maps have a spatial sampling of $\sim$0.5\,kpc at $z\sim1$
 ($0\farcs 06$ pixels).
The data presented in this paper are based on the $\sim70$ pointings,
roughly half of the full data set, that were obtained prior to June
2011 see {Nelson} {et~al.} (2012).
The survey, data reduction, and determination of derived quantities
 (e.g. redshifts and masses) are described in Brammer {et~al.} (2012) 
 and {van Dokkum} {et~al.} (2011).

The 3D-HST grism spectra have high spatial resolution and 
low spectral resolution ($R\sim130$) meaning that emission line structure 
reflects almost exclusively spatial structure (morphology) in 
contrast to data with high spectral resolution
where structure reflects velocity (rotation or dispersion). 
Emission line maps of galaxies are made by subtracting the continuum
emission from the two-dimensional spectrum. The details
of the procedure are described in ({Nelson} {et~al.} 2012; Lundgren {et~al.} 2012;
 Schmidt et al. 2012 submitted) 

We selected galaxies with redshifts $0.8<z<1.3$,
total F140W AB magnitude $<21.9$, and rest-frame \ew\,$>100$\,\AA. 
We also limited the sample 
to galaxies with effectively no contaminating flux from the spectra of other nearby objects.
The position of the sample in the $0.8<z<1.3$ star formation rate 
(SFR)--stellar mass plane is shown in Fig.\,1.
Star formation rates were derived from the IR + UV luminosities
as described in {Whitaker} {et~al.} (2012).
The selected galaxies lie predominantly on the upper half of the 
`star-forming main sequence' 
(as shown in the histogram of Fig 1) 
(e.g.  {Noeske} {et~al.} 2007; {Daddi} {et~al.} 2007; {Whitaker} {et~al.} 2012).
These galaxies were selected in \ha\, not IR; 
while this selection may bias the sample,
it may possibly also have the advantage that it adds less uncertainty to 
interpreting the spatial distribution of star
formation due to dust extinction.

\section{The spatial distribution of star formation}

\subsection{Stacked H$\alpha$ and Stellar Continuum Emission}

\begin{figure*}
\centering
\includegraphics[width=\textwidth]{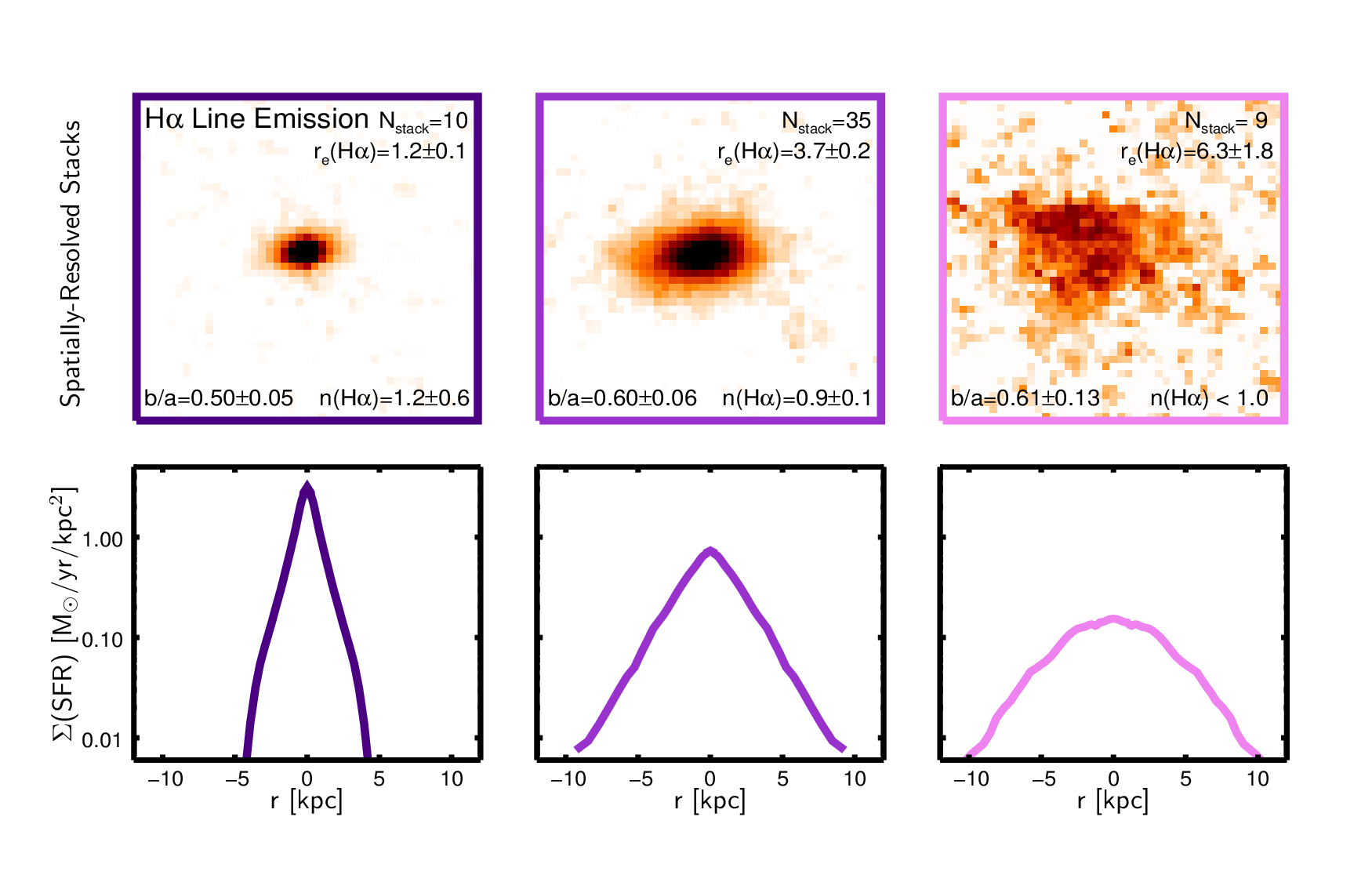} 
\caption{Stacks of the high spatial resolution, PSF-corrected maps of 
 \ha\, emission (top). Stacking was done based on \ha\, size;
 the number of galaxies included in each size bin is listed ($N_{stack}$)
as is the major axis effective radius (in kpc) of each stack measured by galfit (r$_e$). 
The stacks have a weighted mean S{\'e}rsic index $n(\textrm{H}\alpha)\sim1$
and axis ratio of b/a(\ha\,)=0.58$\pm0.09$,
 consistent with disks at random orientation angles. 
Bottom panels show corresponding radial profiles with dark purple -- small,
medium purple -- mid-sized, light purple -- large (Fig.\,3, \S3.2).
The stacked \ha\ emission always has 
 S{\'e}rsic index $n\lesssim1$.
\label{stacks}}
\end{figure*}

We stack the high spatial resolution \ha\, maps from 3D-HST to create
average \ha\, maps -- increasing the S/N
 and providing for a reliable measurement of the 
structural properties of \ha\, to large radii for this sample at $z\sim1$.
We divide the sample in three bins based on their
H$\alpha$ size\footnote{The sizes are measured using growth curves as described
in {Nelson} {et~al.} (2012)} in order
to homogenize the sample and investigate trends with size.\footnote{The
results are similar when the bins are defined according to
the F140W sizes.}
The median stellar masses of the stacks are $\log(\textrm{M}_*)=10.0,10.2,10.2$.
The galaxies are normalized by their F140W flux and centered according
to their luminosity-weighted F140W image centers.
We masked the [S\,{\sc\,ii}] $\lambda\lambda 6716,6731$\AA\, emission in the fit.
At our spectral resolution, the FWHM 
of a line is $\sim100$\,\AA\,,  meaning that \ha\ $\lambda6563$\AA\, and [N\,{\sc\,ii}]
$\lambda6583$\AA\, are unresolved  but \ha\, and [S\,{\sc\,ii}]
are separated by $\sim3$ resolution elements.
We rotated the images based on their GALFIT-derived F140W position angles
to align them along their major axes, summed them, and divided the resulting stack
by the summed masks.
Finally, in order to correct for the effects of the PSF, the stacks
were deconvolved by adding the
GALFIT-derived PSF-deconvolved model to the residuals of the fit, a method 
developed in Szomoru {et~al.} (2010).

Figs.\,2 and 3 show the stacked, PSF-corrected \ha\,
emission. The S/N in the stacks is high, and the emission is clearly
spatially extended. Furthermore, both the rest-frame R-band and \ha\,
maps are clearly elongated. We  quantified the structural properties
of the stacks with
GALFIT (Peng {et~al.} 2002) using
empirical PSFs. For the continuum and H$\alpha$
we find  weighted mean axis ratios of $b/a(R)=0.55\pm0.07$
and $b/a$(\ha\,)=0.58$\pm0.09$ respectively, consistent with 
expectations for randomly oriented disks with thickness c/a=0.26,0.33. 
This thickness could be attributed to disks being intrinsically thick
at $z\sim1$, heterogeneous properties of the galaxies being stacked,
the effect of bulges, a selection biased toward face-on galaxies, or noise.

\subsection{Radial Profiles are Nearly Exponential}

\begin{figure*}
\centering
\includegraphics[angle=270,trim=0mm 0mm 30mm 25mm,clip,width=\textwidth]{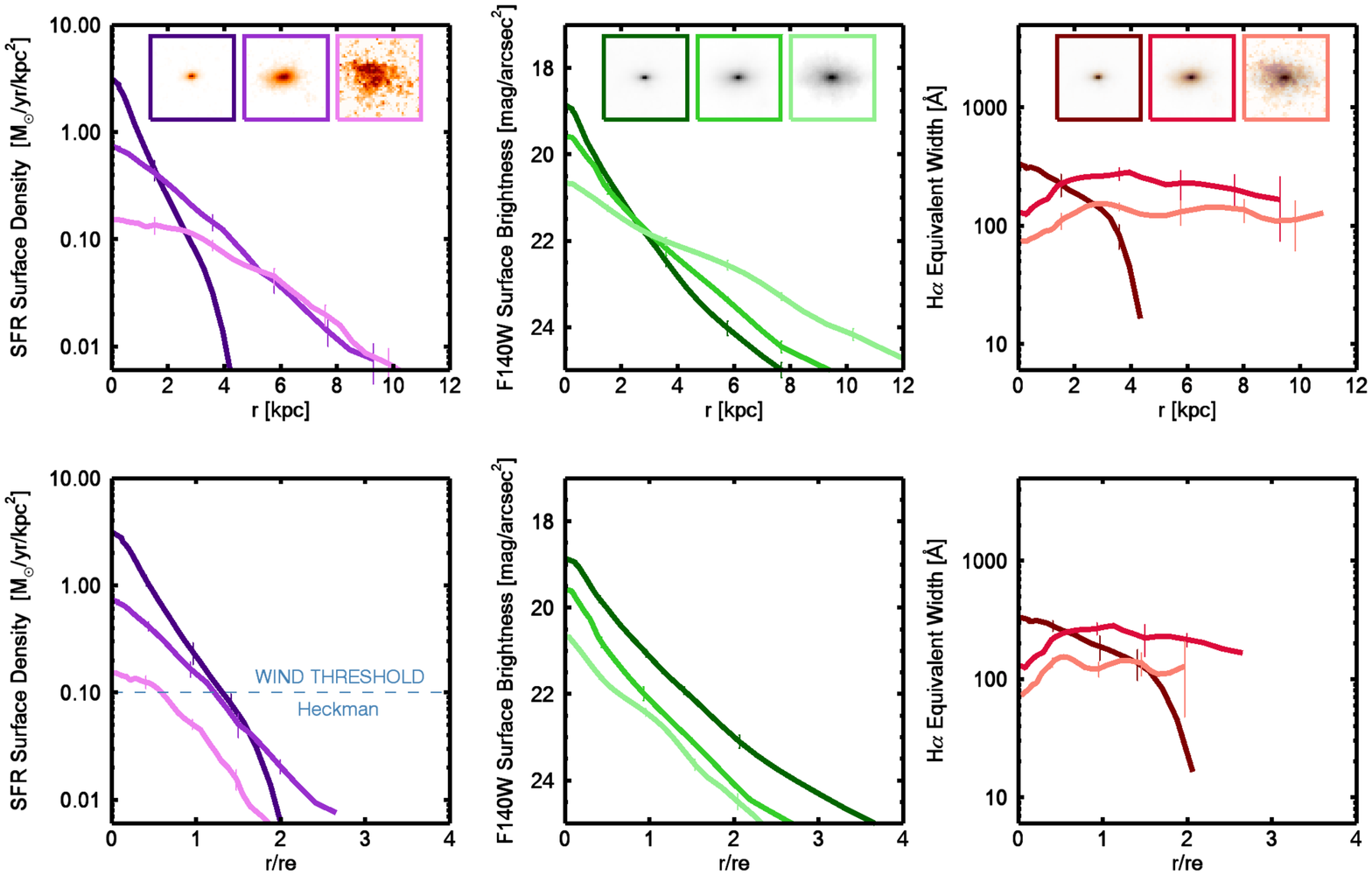}
\caption{The stacks of \ha\, (left) and rest-frame R-band (middle) emission have
nearly exponential (or shallower) radial profiles. 
The EW(\ha\,) profile (discussed in \S3.2) is shown in the right panel.
The bottom panels show the profiles normalized by their effective radius.
The horizontal line is the SFR surface density criterion for driving 
large-scale outflows ({Heckman}~2002)
As in Fig.\,2, the dark, medium, and light colors correspond to the
small, medium, and large stacks respectively as shown at the top.
The spatial distribution of \ha\ emission is exponential or shallower. 
\label{radprof}
}
\end{figure*}

Fig.\,3 shows the radial profiles derived from the stacks.
Radial profiles of the PSF-corrected stacks were created
by summing the flux in elliptical radii. 
A 30\% correction was applied to account for [N\,{\sc\,ii}]
(see \S\,4).
All uncertainties were 
derived by bootstrap resampling the stacks.

The \ha\, emission ranges from 
 exponential to much less centrally concentrated than exponential.
 The stellar continuum emission in all the stacks is
 nearly exponential. 
 
 The radial equivalent width profiles (right panels of Fig.\,3) 
 are flat to within a factor of two and EW(\ha\,)$>100$\AA\, for all stacks
 within 1.5$r_e$. This suggests that the high global EW(\ha\,)s
 seen in these galaxies are not driven primarily by central starbursts
 but by strong star formation at all radii. 
 
The derived
S{\'e}rsic indices (Fig.\,2) of the \ha\, and stellar continuum profiles are 
 $n(\textrm{H}\alpha)=1.2\pm0.6,0.9\pm0.1,\,n(\textrm{H}\alpha)<1$ and 
 $n(\textrm{F140W})=1.8\pm0.4,1.4\pm0.1,1.1\pm0.2$ 
  with weighted means of 
 $n(\textrm{H}\alpha)=1.0\pm0.4$ and $n(\textrm{F140W})=1.4\pm0.2$.
 As shown in {van Dokkum} {et~al.} (2010), structural parameters
 measured from stacks are close to the mean values
 for the individual galaxies going into the stacks.
 The upper limit of $n(\textrm{H}\alpha)<1$ for the largest \ha\, stack 
 reflects the fact that the derived $n(\textrm{H}\alpha)$ depends 
 somewhat on the details of the fit (treatment of sky, fitting region)
 but always $n(\textrm{H}\alpha)<1$.

We find that
the stacked spatial distribution of \ha\, for galaxies in this sample is
exponential or shallower. 
All stacks have $n<2$, implying a bulge fraction of less than
20\% (van Dokkum {et~al.} 1998). 
The radial \ha\,  profiles (Fig.\,3, left),
 seem to show a trend toward lower $n$(\ha\,) with increasing size.
However, this trend is not statistically significant given the errors in the 
derived S{\'e}rsic indices. 
Additionally, note that the median galaxy with EW(\ha\,)$>100$\,\AA\ 
has a high enough star formation surface density to drive
 winds out to $\sim1\,r_e$ ({Heckman}~2002).

\subsection{Effects of Dust}

\begin{figure*}
\centering
\includegraphics[width=0.9\textwidth]{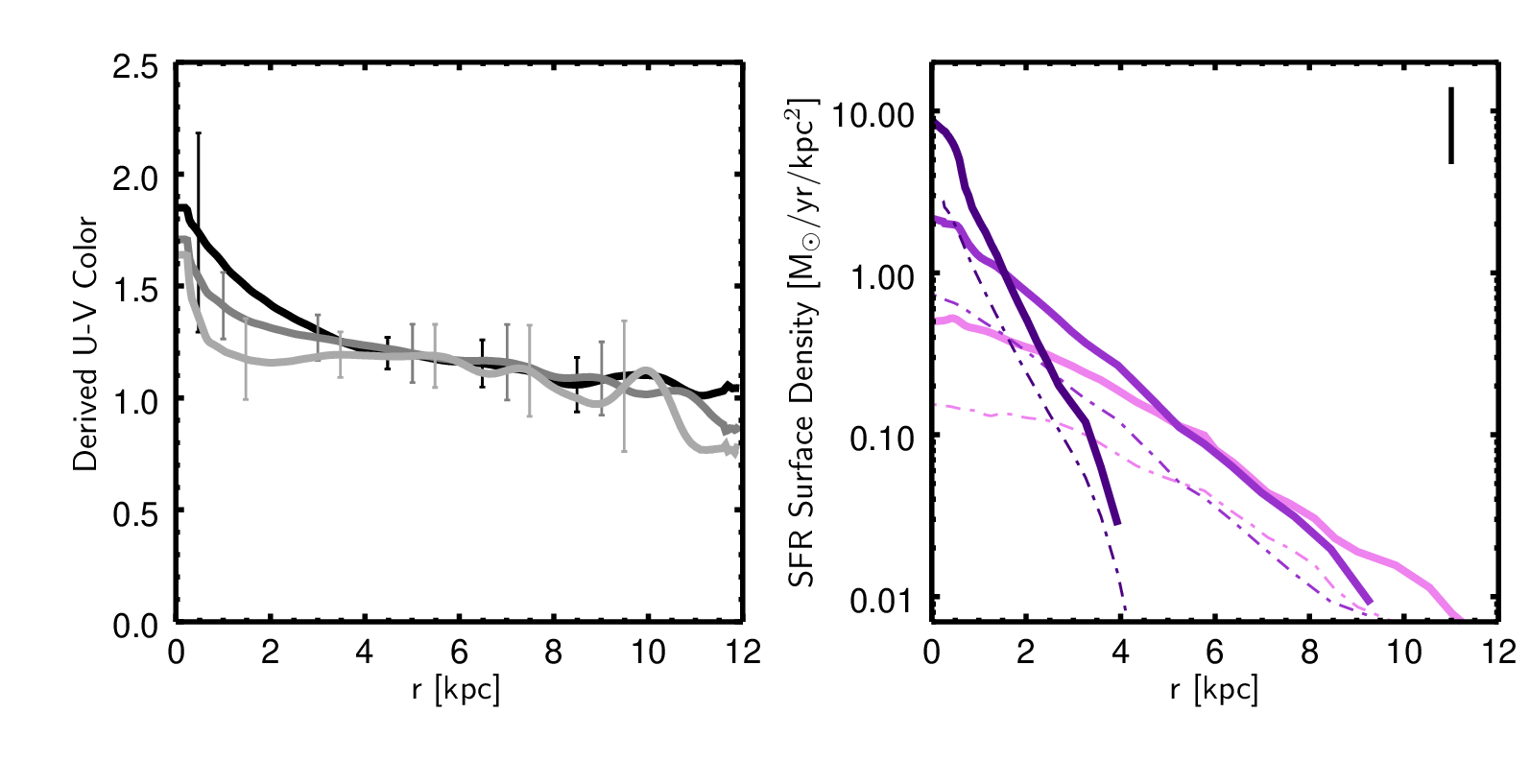}
\caption{The left panel shows the radial color profiles for each of the stacks
with black for the small stack, dark gray for the medium stack, and light gray 
for the large stack. The right panel shows the implied dust corrected
radial profiles of star formation surface density with the colors as in 
Fig. 2 and 3. Error bar on the right panel denotes a typical uncertainty.
Dust-corrected radial profiles of star formation remain nearly exponential.
\label{dust}
}
\end{figure*}

A major uncertainty in the interpretation of the radial \ha\, profile is the effect of
differential dust extinction: if some parts of the galaxies are more
obscured than others (see e.g.  {Wuyts} {et~al.} 2012),
 the derived radial profile of \ha\, would not
reflect the radial profile of star formation. 
To assess the importance
of this effect, we estimate the extinction as a function of radius.
We determine the extinction from the radial color profile. 

The color profile is determined from the 
combination of  the F140W stacks with 
ACS F814W stacks, which can be converted into rest-frame $U-V$ color
profiles. As shown in the left panel of Fig.\,4,
for most of the radial extent of the stacks the color is 
fairly constant, but it becomes redder
inside a radius of 2\,kpc  (see also Szomoru {et~al.} 2012).
To estimate roughly
how color translates into missed star formation,
we assume that the star formation that is not captured by \ha\, will be captured 
by the IR (Kennicutt {et~al.} 2009). 
Using photometry from Skelton et al. (2013) in prep, in addition to 
 rest-frame U-V colors and IR-based SFRs from 
 the NEWFIRM Medium Band Survey  (NMBS,  {Whitaker} {et~al.} 2012),
empirical relations were derived for the translation of
rest-frame U-V colors into star formation corrections:
\begin{equation}
\log(\textrm{SFR(IR+H}\alpha))-\log(\textrm{SFR(H}\alpha)) \sim 0.3\times(U-V)
\end{equation}

The dust-corrected radial profiles
are shown in the right panel of Fig 4. The smallest
galaxies appear to have star formation which is marginally steeper
than exponential. The large galaxies have the largest radial gradients
but their implied star formation is still less steep than exponential.
Although this analysis assigns the entire observed color gradient
to extinction, the centers of the radial color profiles could be
red because of dust or age. 
Importantly, the S{\'e}rsic indices of the star formation in these stacks 
based on the implied dust correction remain close to one -- $n=1.4,1.1,0.6$ 
for each of the stacks, weighted mean=1.0$\pm 0.4$
So, even when accounting for dust, the averaged star formation in
these galaxies has a nearly exponential spatial distribution.


\section{Kinematics}
\begin{figure*}
\centering
\includegraphics[trim=0 70mm 20mm 0,clip,width=\textwidth]{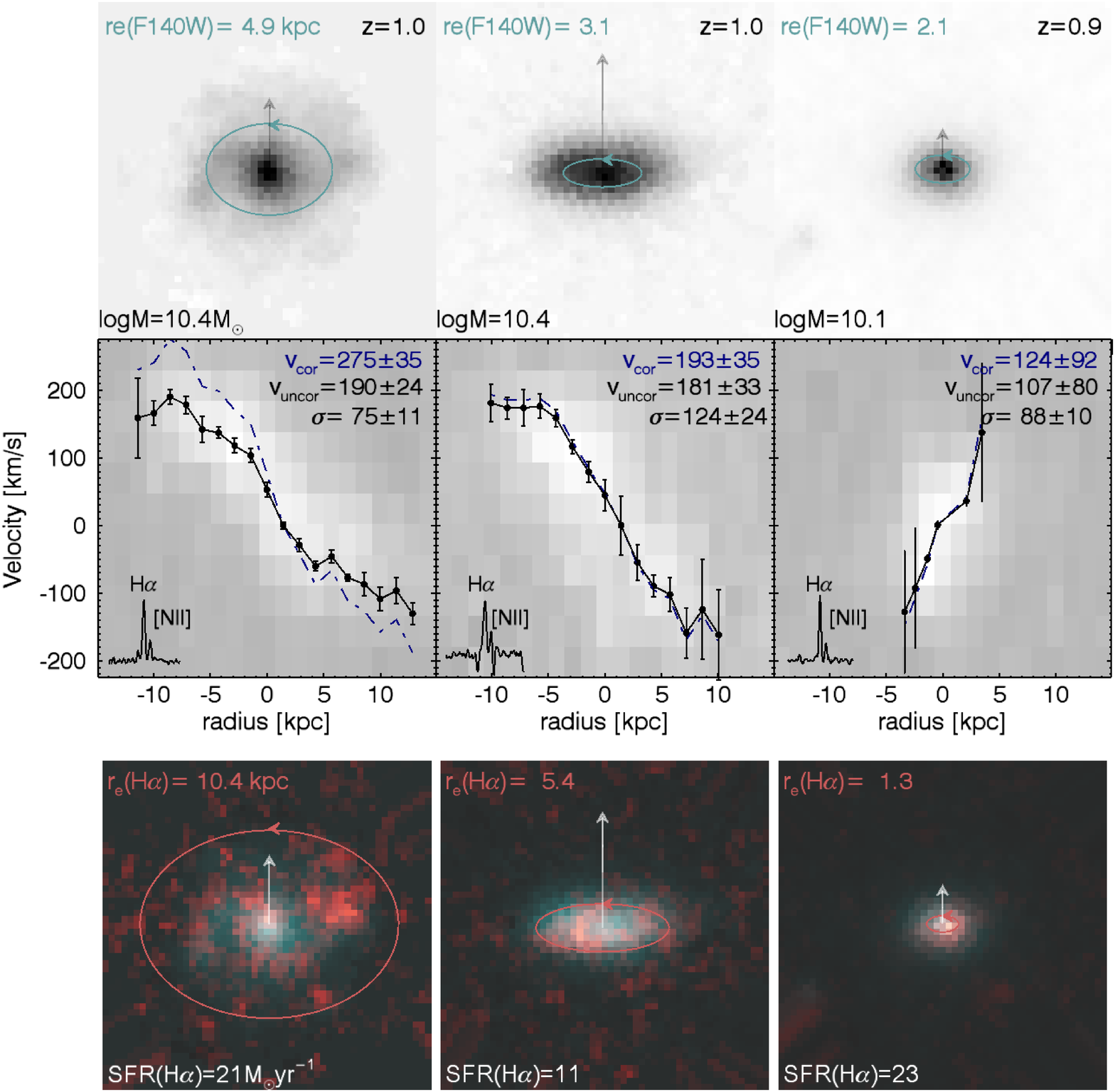}
\caption{Shown here are the rotation curves (middle row) 
derived from the NIRSPEC spectra (insets in corners),
 one example from each stack with the implied rotation velocities 
 (v$_{\textrm{corr}}$ -- corrected and v$_{\textrm{uncorr}}$ -- uncorrected for inclination)
 and velocity dispersion ($\sigma$) listed in km/s. 
The top row shows the corresponding rest-frame R-band images
and the bottom row shows a false color image with the \ha\, emission in red 
and stellar continuum in blue. 
In these rows, ellipses mark the R-band and \ha\, effective radii 
($r_e$(R), $r_e$(\ha\,)) respectively 
and gray arrows have a scaled length corresponding to the projected 
specific angular momentum. 
\label{kinematics}
}
\end{figure*}

The flattening of the stacks
and the exponential H$\alpha$ profiles suggest that the star formation
occurs in rotating disks. To test this,
we measured kinematics for a subset of this
sample using NIRSPEC 
on the Keck II telescope on April 9-10, 2012. 
The sample comprises eight galaxies 
chosen from the small~(1), middle~(5), and large~(2) stack,
 which have sizes, masses, S{\'e}rsic indices, star formation rates,
and axis ratios representative of the sample as a whole. 
 
We used the low dispersion mode of NIRSPEC
with 0.5" seeing and a slit width of 0.7", giving a spectral resolution of
$\sigma\sim80$km/s in the $J$ band
 (compared to $\sim500$km/s in the grism spectra).
The slit was aligned along the major axis of 
each galaxy and observations were conducted 
in a series of four 900s exposures, dithering along the slit.
The data reduction followed standard procedures for long
slit spectroscopy (see, e.g.,  van Dokkum {et~al.} 2004).
We extracted kinematic information 
from the two dimensional spectra by fitting a gaussian to 
the \ha\, and [N\,{\sc ii}] emission simultaneously at each spatial position
along the slit.
The median [N\,{\sc ii}]/\ha\, is 0.3.
Assuming the velocity shear in the two dimensional spectra is
due to rotation, we take the rotational velocity to be the velocity
difference between the geometrical center of the galaxy 
 and the maximum velocity.
We correct the rotation velocities
for inclination angle using the 3D-HST axis ratios. 
We calculate one dimensional velocity dispersions by fitting
a gaussian to the \ha\ emission in the central row of each
spectrum. We calculate the instrumental $\sigma$ to be 
$\sim 80$km/s by fitting a gaussian to the sky lines.
The intrinsic velocity dispersion is calculated by subtracting 
the instrumental dispersion from the measured dispersion in 
quadrature. 

We find that all galaxies show velocity shear in their 
two dimensional spectra, with derived rotation 
velocities of 90--235 km/s uncorrected and 
110--330 km/s corrected for inclination.\footnote{Measured velocity dispersions
are also relatively high, but are difficult to interpret given the
low spatial resolution of the Keck spectra.}
Fig.\,5 shows the rotation curves for one example galaxy
for each stack. 
These spectra are of the best quality but appear to be representative
of the eight.
The velocity profile of the smallest galaxy in Fig.\,3 does not show evidence
  for a turnover, which means the measured maximum velocity
  is a lower limit on the rotation velocity at large radii.
We conclude that the kinematics are consistent with the disk interpretation
of the structural properties of the \ha\, emission.
Although different classification methods make it difficult to
perform a quantitative comparison, our results
are qualitatively consistent with the finding that a large
fraction of star-forming galaxies at $z\sim1$
appear to be rotating.
(e.g.,  Wisnioski {et~al.} 2011; 
Epinat {et~al.} 2012; Swinbank {et~al.} 2012).

\section{Discussion}

The central result of this paper is that the radial distribution of
stacked \ha\, emission in $z\sim 1$ galaxies is close to
exponential out to $\sim 10$ kpc. Combined with the axis ratios of
the stacks and the kinematics of a subset of the sample, the most
straightforward interpretation is that star formation seems typically
to occur in disks at $z\sim1$ at least for galaxies with high EW(\ha\,).
The factor of $\sim10$
increase in the star formation rates of galaxies from $z=0$ to $z=1$
(e.g.,  {Damen} {et~al.} 2009; Fumagalli {et~al.} 2012) 
is apparently driven by increased star formation
activity in disks rather than a much greater prevalence of merger-driven
central starbursts -- consistent with other studies
 (e.g.,  {Rodighiero} {et~al.} 2011; {Wuyts} {et~al.} 2011, 2012).

This result raises a number of questions.
First, it is not clear how these galaxies are related to
$z\sim0$ galaxies. 
An average galaxy in this sample has 
a distribution of stars with a S{\'e}rsic index of $n=1.4$.
If this average galaxy forms stars with $n=1.0$,
it will have a lower S{\'e}rsic index at later times.
 Either we are witnessing the build up of only the
 latest of late-type galaxies; these galaxies substantially 
 change where their star formation is occurring between $z\sim1$ and $z\sim0$; 
 or their stars, once formed, must migrate into a different 
 configuration. In other words, if these are the ancestors 
 of typical $z\sim0$ spiral galaxies, their bulges need to be built 
 in some way besides the star formation we are seeing.
 This could be accomplished by mergers or secular evolution
 (e.g., Bournaud, Elmegreen, \& Martig 2009; {Ellison} {et~al.} 2011).
These observations could be understood in the context of the following picture:
at $z>1$, disks are gas-rich and turbulent.
 As redshift decreases, the gas fraction decreases, 
 and a bar instability grows, increasing the central concentration 
 and central vertical velocity dispersion. 
(e.g., van~den~Bosch 2001; {DeBuhr}, {Ma}, \& {White} 2012;
 {Forbes}, {Krumholz}, \&  {Burkert} 2012) 

We also note the somewhat surprising presence of a
population of large, rapidly-rotating disks with relatively
low stellar masses at $z\sim1$. 
These galaxies have mean $r_e$(\ha)=7.3\,kpc, v=240\,km/s, $\sigma=89$\,km/s,
and $M_*=2.2\times10^{10}$\msun, meaning they are disks 8\,Gyr ago,
 larger in size than the Milky Way (Drimmel \& Spergel 2001) and thicker ({Freeman} 1987),
 with $\sim1/3$ of the stellar mass ({McMillan} 2011) and
  similar or higher maximum rotation velocities (Bovy, Hogg, \& Rix 2009).
Both what these galaxies become in the local universe 
and how such extended star forming disks were made in an epoch
of high disk turbulence are open questions.
The small disks on the other end of the size distribution are also
of interest.
With median $r_e$(\ha)=1.9\,kpc, $r_e$(R)=2.1\,kpc, 
 $M_*=8.9\times10^9$M$_{\odot}$, and stacked $n(\textrm{H}\alpha)=1.2$,
they more closely resemble compact versions of disky 
star formers than merger-driven star formation in spheroids
(see also van~der~Wel {et~al.} (2011).)

There are several caveats. First, our sample is not complete in mass
or in SFR(UV+IR). In future papers (using the full 3D-HST
survey) we will study the distribution of star formation as a function of
these parameters. Second, the distribution of star formation in the stacks
is  not necessarily representative of individual galaxies. In individual
galaxies the  star formation is clumpy
 (e.g. {Genzel} {et~al.} 2008, 2011; {F{\"o}rster Schreiber} {et~al.} 2009; {Nelson} {et~al.} 2012, Fig.\,1)
and it is difficult to quantify the structure. Stacking these clumpy objects
could be hazardous, but it may also provide more insight than can be gleaned
from individual galaxies: assuming that the clumps
are transient and ``light up'' a part of the underlying gas disk
for a short time (as in e.g. {Wuyts} {et~al.} 2012), our stacking technique
effectively produces a time-averaged map of the star formation in the
galaxies. Finally, dust attenuation remains a key uncertainty, both in
the selection and in the interpretation of the profiles.
This can be addressed with maps of the IR emission
at the full ALMA resolution, or
by measuring (spatially-resolved) Balmer decrements.

We thank the referee for a thorough report which improved the paper.
Support from grant HST GO-12177 is gratefully acknowledged.

.


\begin{references}

\reference{} Bournaud, F., Elmegreen, B.~G., \& Martig, M. 2009, \apjl,  707, 1

\reference{} Bovy, J., Hogg, D.~W., \& Rix, H.-W. 2009, \apj, 704, 1704

\reference{} Brammer, G.~B., {et~al.} 2012, eprint arXiv:1206.1867, 200, 13

\reference{} Brooks, A.~M., Governato, F., Quinn, T., Brook, C.~B., \& Wadsley, J. 2009, The  Astrophysical Journal, 694, 396

\reference{} Cresci, G., {et~al.} 2009, The Astrophysical Journal, 697, 115

\reference{} {Daddi}, E., {et~al.} 2007, \apj, 670, 156

\reference{} {Damen}, M., {Labb{\'e}}, I., {Franx}, M., {van Dokkum}, P.~G., {Taylor},  E.~N., \& {Gawiser}, E.~J. 2009, \apj, 690, 937

\reference{} {DeBuhr}, J., {Ma}, C.-P., \& {White}, S.~D.~M. 2012, \mnras, 426, 938

\reference{} {Dekel}, A., {Sari}, R., \& {Ceverino}, D. 2009, \apj, 703, 785

\reference{} Drimmel, R., \& Spergel, D.~N. 2001, The Astrophysical Journal, 556, 181

\reference{} {Ellison}, S.~L., {Patton}, D.~R., {Nair}, P., {Simard}, L., {Mendel}, J.~T.,  {McConnachie}, A.~W., \& {Scudder}, J.~M. 2011, in IAU Symposium, Vol. 277,  IAU Symposium, ed. C.~{Carignan}, F.~{Combes}, \& K.~C. {Freeman}, 178--181

\reference{} Epinat, B., {et~al.} 2009, Astronomy and Astrophysics, 504, 789

\reference{} Epinat, B., {et~al.} 2012, Astronomy and Astrophysics, 539, A92

\reference{} {Forbes}, J., {Krumholz}, M., \& {Burkert}, A. 2012, \apj, 754, 48

\reference{} {F{\"o}rster Schreiber}, N.~M., {et~al.} 2009, \apj, 706, 1364

\reference{} {F{\"o}rster Schreiber}, N.~M., {Shapley}, A.~E., {Erb}, D.~K., {Genzel}, R.,  {Steidel}, C.~C., {Bouch{\'e}}, N., {Cresci}, G., \& {Davies}, R. 2011, \apj,  731, 65

\reference{} {Freeman}, K.~C. 1987, \araa, 25, 603

\reference{} Fumagalli, M., {et~al.} 2012, eprint arXiv:1206.1867, 757, L22

\reference{} {Genzel}, R., {et~al.} 2008, \apj, 687, 59

\reference{} {Genzel}, R., {et~al.} 2011, \apj, 733, 101

\reference{} Grand, R. J.~J., Kawata, D., \& Cropper, M. 2012, eprint arXiv:1206.1867, 421,  1529

\reference{} Heckman, T. M. 2002, Extragalactic Gas at Low Redshift, 254, 292

\reference{} {Hopkins}, P.~F., {Hernquist}, L., {Cox}, T.~J., {Di Matteo}, T., {Robertson},  B., \& {Springel}, V. 2006, \apjs, 163, 1

\reference{} Jones, T.~A., Swinbank, A.~M., Ellis, R.~S., Richard, J., \& Stark, D.~P. 2010,  \mnras, 404, 1247

\reference{} Kassin, S.~A., Weiner, B., Faber, S. M., {et~al.} 2012, \apj, 758, 106

\reference{} Kennicutt, R.~C., {et~al.} 2009, The Astrophysical Journal, 703, 1672

\reference{} {Law}, D.~R., {Steidel}, C.~C., {Erb}, D.~K., {Larkin}, J.~E., {Pettini}, M.,  {Shapley}, A.~E., \& {Wright}, S.~A. 2009, \apj, 697, 2057

\reference{} Lundgren, B.~F., {et~al.} 2012, \apjl, 760, 49

\reference{} Mancini, C., {et~al.} 2011, The Astrophysical Journal, 743, 86

\reference{} {McLean}, I.~S., {et~al.} 1998, in Society of Photo-Optical Instrumentation  Engineers (SPIE) Conference Series, Vol. 3354, Society of Photo-Optical  Instrumentation Engineers (SPIE) Conference Series, ed. A.~M. {Fowler},  566--578

\reference{} {McMillan}, P.~J. 2011, \mnras, 414, 2446

\reference{} {Nelson}, E.~J., {et~al.} 2012, \apjl, 747, L28

\reference{} {Noeske}, K.~G., {et~al.} 2007, \apjl, 660, L43

\reference{} Peng, C.~Y., Ho, L.~C., Impey, C.~D., \& Rix, H.-W. 2002, The Astronomical  Journal, 124, 266

\reference{} {Rodighiero}, G., {et~al.} 2011, \apjl, 739, L40

\reference{} Ro{\v s}kar, R., Debattista, V.~P., Quinn, T.~R., Stinson, G.~S., \& Wadsley,  J. 2008, The Astrophysical Journal, 684, L79

\reference{} {Shapiro}, K.~L., {et~al.} 2008, \apj, 682, 231

\reference{} Swinbank, M., Smail, I., Sobral, D., Theuns, T., Best, P., \& Geach, J. 2012,  arXiv: 1209:1395

\reference{} Szomoru, D., {et~al.} 2010, eprint arXiv:1206.1867, 714, L244

\reference{} Szomoru, D., Franx, M., van Dokkum, P.~G., Trenti, M., Illingworth, G.~D.,  Labbe, I., \& Oesch, P. 2012, arXiv: 1208:4363

\reference{} van den~Bosch, F.~C. 2001, Monthly Notices of the Royal Astronomical Society,  327, 1334

\reference{} van~der Wel, A., {et~al.} 2011, The Astrophysical Journal, 742, 111

\reference{} van Dokkum, P.~G., Franx, M., Kelson, D.~D., Illingworth, G.~D., Fisher, D., \&  Fabricant, D. 1998, The Astrophysical Journal, 500, 714

\reference{} van Dokkum, P.~G., {et~al.} 2004, The Astrophysical Journal, 611, 703

\reference{} {van Dokkum}, P.~G., {et~al.} 2010, \apj, 709, 1018

\reference{} {van Dokkum}, P.~G., {et~al.} 2011, \apjl, 743, L15

\reference{} {Whitaker}, K.~E., {van Dokkum}, P.~G., {Brammer}, G., \& {Franx}, M. 2012,  \apjl, 754, L29

\reference{} Wisnioski, E., {et~al.} 2011, Monthly Notices of the Royal Astronomical  Society, 417, 2601

\reference{} {Wuyts}, S., {et~al.} 2012, \apj, 753, 114

\reference{} {Wuyts}, S., {et~al.} 2011, \apj, 742, 96

\end{references}
\end{document}